# A Tunable Ferroelectric Based Unreleased RF Resonator

Yanbo He[*], he355@purdue.edu, Purdue University, West Lafayette, IN, USA

Bichoy Bahr, bichoy@ti.com, Kilby Labs, Texas Instruments, Dallas, TX, USA

Mengwei Si, msi@purdue.edu, Purdue University, West Lafayette, IN, USA

Peide Ye, yep@purdue.edu, Purdue University, West Lafayette, IN, USA

Dana Weinstein, danaw@purdue.edu, Purdue University, West Lafayette, IN, USA

*Address: 1205 W State St, West Lafayette, IN 47907

t. 765.496.3647

e. he355@purdue.edu

## ABSTRACT

This paper introduces the first tunable ferroelectric capacitor (FeCAP)-based unreleased RF MEMS resonator, integrated seamlessly in Texas Instruments' 130nm Ferroelectric RAM (FeRAM) technology. An array of FeCAPs in this complementary metal-oxide-semiconductor (CMOS) technology's back-end-of-line (BEOL) process were used to define the acoustic resonance cavity as well as the electromechanical transducers. To achieve high quality factor ($Q$) of the resonator, acoustic waveguiding for vertical confinement within the CMOS stack is studied and optimized. Additional design considerations are discussed to obtain lateral confinement and suppression of spurious modes. An FeCAP resonator is demonstrated with fundamental resonance at 703 MHz and $Q$ of 1012. This gives a frequency-quality factor product $f \cdot Q = 7.11 \times 10^{11}$ which is 1.6x higher than the most state-of-the-art Pb(Zr,Ti)$O_3$ (PZT) resonators. Due to the ferroelectric characteristics of the FeCAPs, transduction of the resonator can be switched on and off by adjusting the electric polarization. In this case, the resonance can be turned off completely at ±0.3V corresponding to the coercive voltage of the constituent FeCAP transducers. These novel switchable resonators may have promising applications in on-chip timing, ad-hoc radio front ends, and chip-scale sensors.

## INTRODUCTION

Tunable high-Q, small footprint resonators have great potential in both mature and emergent fields such as radio frequency (RF) components[1,2] and communication[3], timing[4,5], sensing[6–8], and imaging[9]. The demanding performance of these devices and systems, alongside requirements for miniaturization, lower power consumption, and lower cost, has pushed the limits on what conventional technology can achieve[10].

The authors have previously demonstrated high-Q, unreleased resonators with field effect transistor (FET) electromechanical sensing, referred to as Resonant Body Transistors (RBTs)[11] in standard CMOS technology at frequencies ranging from 3 GHz[12] to 32 GHz[13]. While the $f \cdot Q$ products of these devices are record breaking, their return loss and bandwidth are restricted by the fundamental limits of electrostatic transduction, which provides modest driving force density. This is particularly evident in the case of planar CMOS technology (e.g. 32nm SOI), where the electromechanical transconductance of a 3 GHz resonator is on the order of 100nS[14]. The correspondingly high insertion loss (IL) of such a device makes oscillator and filter design challenging. It is therefore necessary to explore alternative IC integrated materials.

Ferroelectric materials have spontaneous electrical polarization which can be reoriented under an external electric field[15,16]. The hysteretic behavior of ferroelectrics has driven fundamental investigation[17,18] and commercial development for integrated circuit (IC) memory and has sparked a class of devices termed ferroelectric RAM (FeRAM or FRAM)[19,20]. Ferroelectrics have an additional trait of piezoelectric response; the changing polarization within the ferroelectric material induces dielectric dipole moment and changes the lattice constant, generating internal stress in the material[21]. Commonly used ferroelectric materials include barium titanate ($BaTiO_3$)[22], barium strontium titanate (BST)[23,24], lead zirconium titanate (PZT)[17,25], lead titanate ($PbTiO_3$)[26] and hafnium dioxide ($HfO_2$) [27,28].

Texas Instruments (TI) E035 FeRAM technology has integrated ferroelectric PZT in the back-end-of-line (BEOL) process of their 130nm CMOS technology[29]. Leveraging this IC platform, we can realize ferroelectric based CMOS-MEMS resonators with higher electromechanical coupling coefficient ($k^2$) than their electrostatic counterparts. The boost in performance facilitates larger bandwidth filters, lower power oscillators, and higher frequency tolerance to fabrication variations. This paper reports on the first piezoelectric resonators designed in TI's FeRAM process. These devices are the first implementation of unreleased resonators based on ferroelectric capacitors.

## DEVICE DESIGN AND FABRICATION

In order to define the resonance mode with highest $Q$ and transduction efficiency, acoustic waves must be well confined vertically in the CMOS stack with stress maximized in the PZT layer within the ferroelectric capacitors. Analogous to optical waveguide design, this is achieved using acoustic waveguiding within the CMOS stack. Assuming an infinitely long translationally invariant structure, we first determine the dispersion relations of the CMOS stack to define modes restricted to propagation in the plane of the wafer.

The device is first divided into multiple periodic unit cells with lattice constant *a*. To impose lateral (*x*-direction) translational symmetry, the left and right boundaries are defined as Floquet periodic boundary conditions

$$\vec{u}_R(\vec{r}) = \vec{u}_L(\vec{r}) \cdot e^{-i\vec{k}\cdot\vec{r}} \quad (1)$$

where $\vec{u}_L$ and $\vec{u}_R$ are the displacement field at the left and the right boundaries of the single unit cell. $\vec{k}$ is the wave-vector which, due to the periodicity of the lattice, can be determined from the reciprocal lattice with a periodicity of $\pi/a$. This corresponds to the lateral width of the first irreducible Brillouin zone in the reciprocal lattice.

Two-dimensional finite element analysis is then performed in COMSOL Multiphysics® to map these waveguided modes. By applying Perfectly Matched Layers (PML) at the top and bottom of the device, no reflections from these boundaries are considered, approximating an infinitely thick Si substrate and thick, acoustically lossy dielectric layer in the BEOL.

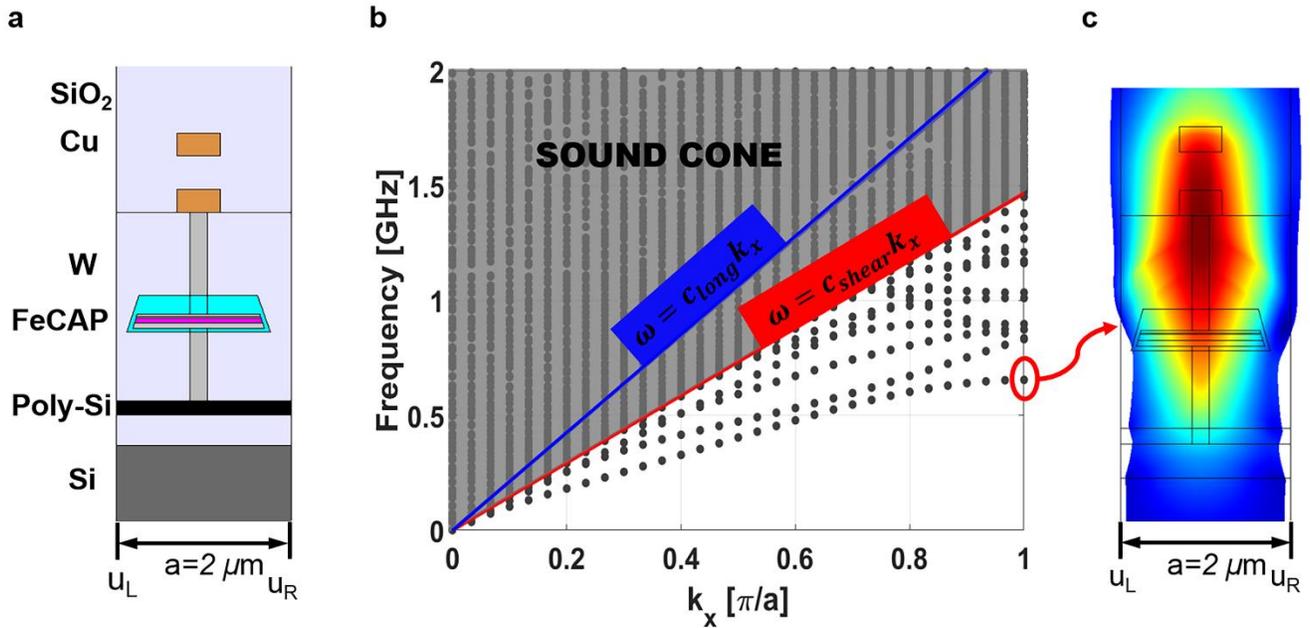

**Fig. 1 a** Schematic of a single unit cell. **b** Dispersion relation showing the vertically confined mode. Modes in the sound cone are free to propagate into the bulk Si substrate and cannot be well confined. **c** Corresponding displacement field of the localized mode shape of the unit cell FeCAP.

Several designs are put forward and optimized. The details of the optimization process are summarized in Supplementary Information. It is found that a device with trapezoidal FeCAP spanning 1.4 *μm* and two 0.6*μm* wide metal layers above it exhibits the best vertical acoustic confinement. The corresponding schematics of a single unit cell is shown in Fig. 1(a). The resulting dispersion relation for the acoustic modes in the CMOS stack is shown in Fig. 1(b). Each point in the dispersion relation corresponds to an eigenmode of the periodic FeCAP structure. This dispersion relation can be divided into three regions separated by sound-lines (red and blue) represented by $\omega = c \cdot k$ where $c$ is the corresponding longitudinal (blue) or shear (red) acoustic velocity in the Si substrate. Above the longitudinal sound-line is a region where all modes are free to propagate, referred to as the "sound cone". Additionally, below the shear sound line exist several discrete modes with sufficiently low acoustic velocity, indicating that the elastic waves are prohibited from propagation in the bulk and the elastic energy is therefore confined in the BEOL region of the CMOS chip. It was previously shown that driving along $k_x=\pi/a$ is beneficial for reducing scattering to the sound cone, enabling higher $Q$[12]. In the meantime, the farther away the activated mode from the sound line, the better confinement the resonator will achieve.

Hence, Fig. 1(c) shows a vertically confined mode at 700 MHz where the strain is well contained in the FeCAP, W via and Cu metal.

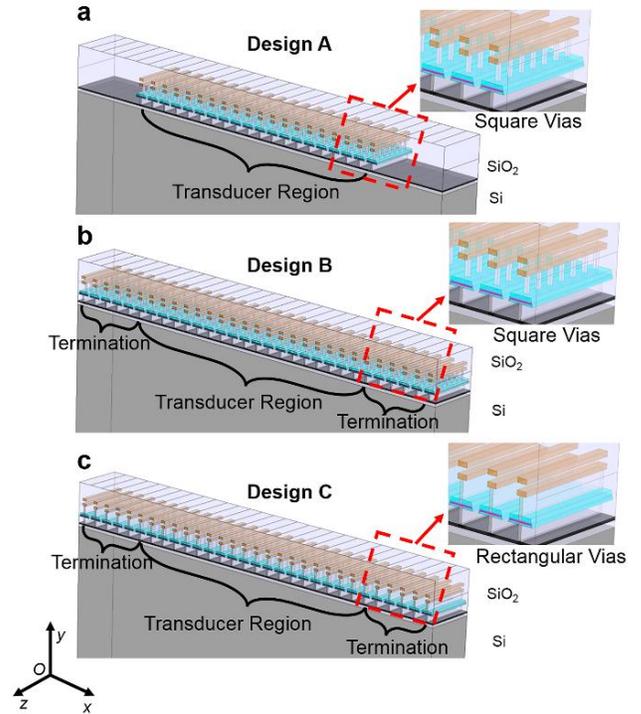

**Fig. 2 Schematics of resonator a** with only main resonant cavity, **b** containing both main cavity and two terminations at the sides, and **c** with main cavity, terminations and traditional square vias replaced by rectangular vias.

To optimize lateral confinement of the acoustic mode, termination of the acoustic waveguide in the plane of the wafer must be carefully considered. In Design A (Fig. 2(a)), the resonator is abruptly terminated at either end of the transducer region, leading to larger impedance mismatch, and subsequent scattering loss. In Design B (Fig. 2(b)), dummy FeCAP unit cells are added to either end of the transducers. These elements are spaced with the same period as the FeCAP transducer region. In this way, the boundaries of the transducers are extended towards the ends and the scattering losses inside the transducer region can be minimized. Furthermore, in traditional CMOS technology, "discrete-block" vias are used to electrically route the FeCAPs to the first metal layer. Thus, to avoid scattering losses along $\vec{oz}$ direction, as shown in Fig. 2(c), traditional discrete vias are replaced with continuous, rectangular, "wall-like" vias.

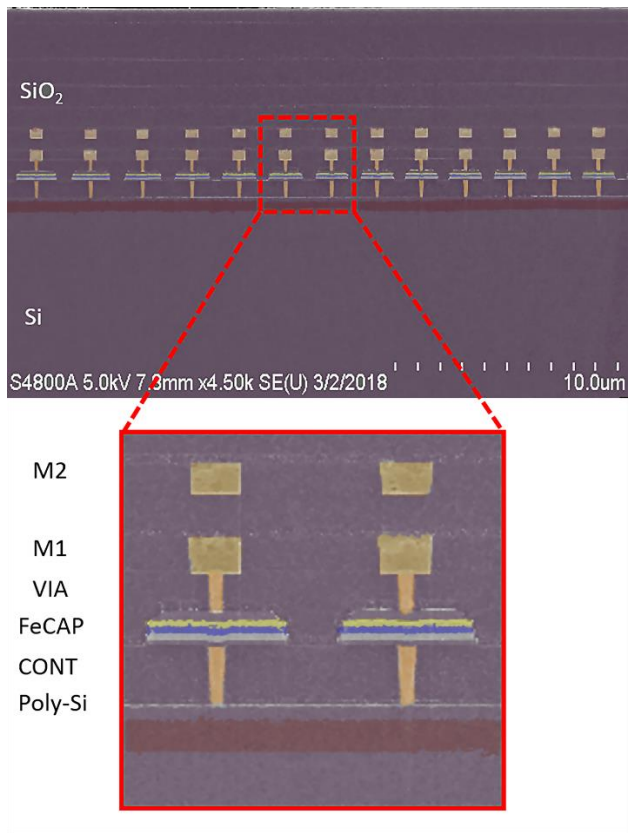

**Fig. 3** Cross sectional scanning electron micrograph of the CMOS-MEMS FeCAP-based resonator[30].

The corresponding scanning electron micrograph (SEM) of a side view schematic of the proposed resonator is plotted in Fig. 3, showing an array of trapezoidal FeCAPs each spanning 1.4 $\mu m$ in length, connected to Cu strips by W vias. The resonator consists of 20 transducers alternating with 2 $\mu m$ periodicity to form the resonant cavity. The resonator has an overall footprint of 108 $\mu m$ by 7 $\mu m$.

## EXPERIMENTAL RESULTS AND DISCUSSION

The ferroelectric properties of the FeCAP were first characterized to investigate PZT film behavior. Polarization-Electric field (P-E) measurement was carried out on a Radiant Technologies RT66C ferroelectric tester, shown in Figure 4(a). The operational voltages for the FeCAPs are limited between -1.5 V and 1.5 V by dielectric leakage. A coercive voltage of ±0.3 V is obtained with remnant polarization of 16.86 $\mu C/cm^2$. The saturation polarization is 38 $\mu C/cm^2$.

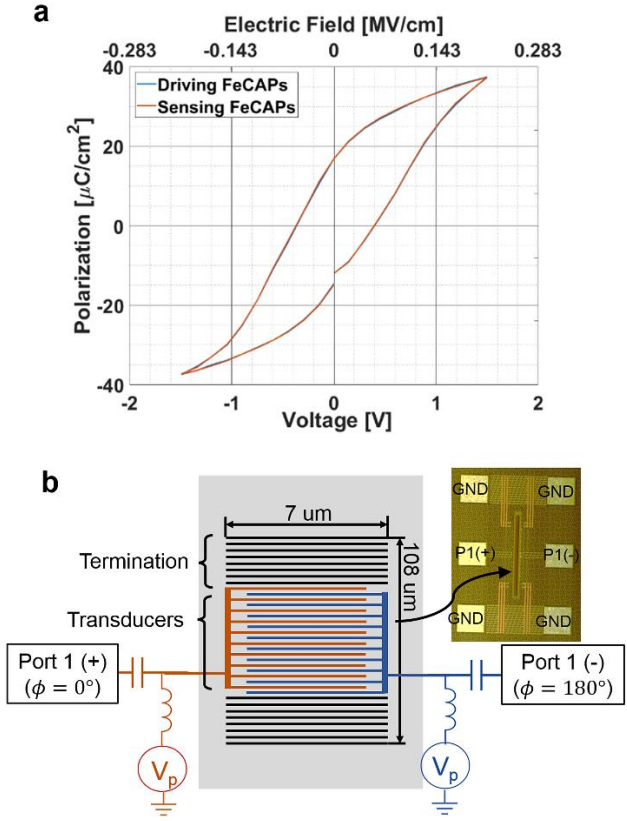

**Fig. 4 a** Measured hysteresis of ferroelectric PZT FeCAP polarization. For efficient piezoelectric transduction, a polarization voltage $V_P$ between -1.5V and 1.5V is applied across the device[30]. **b** Experimental setup to characterize the FeCAP resonator, including a schematic of the port connection to the transducers. Electrically isolated FeCAPs on either side of these transducers (black) are used to terminate the resonance cavity and provide in-plane elastic confinement. Inset shows an optical micrograph of the device.

As previously noted, the entire device consists of 20 FeCAPs transducers. This gives a total FeCAP area of 196 $\mu m^2$ on the transducers. The experimental setup is shown in Figure 4(b).

To enforce the excitation of the desired mode at the Brillouin zone edge ($k = \pi/a$), a differential signal was applied between alternating transducers. Differential 1-port RF measurement was performed in ambient pressure and temperature using a Keysight N5225A PNA. The poling voltage was provided by two Keithley 2400 source measurement units (SMUs), connected to ports 1(+) and 1(−) through bias tees. All transducers are biased at same poling voltage ($V_p$). The RF power is -10 dBm which corresponds to a peak-to-peak voltage of 200 mV. The IF bandwidth (IFBW) is 500 Hz. The differential S-parameters can be obtained by the following equation

$$S_{dd11} = (S_{11} + S_{22} - S_{12} - S_{21})/2 \qquad (2)$$

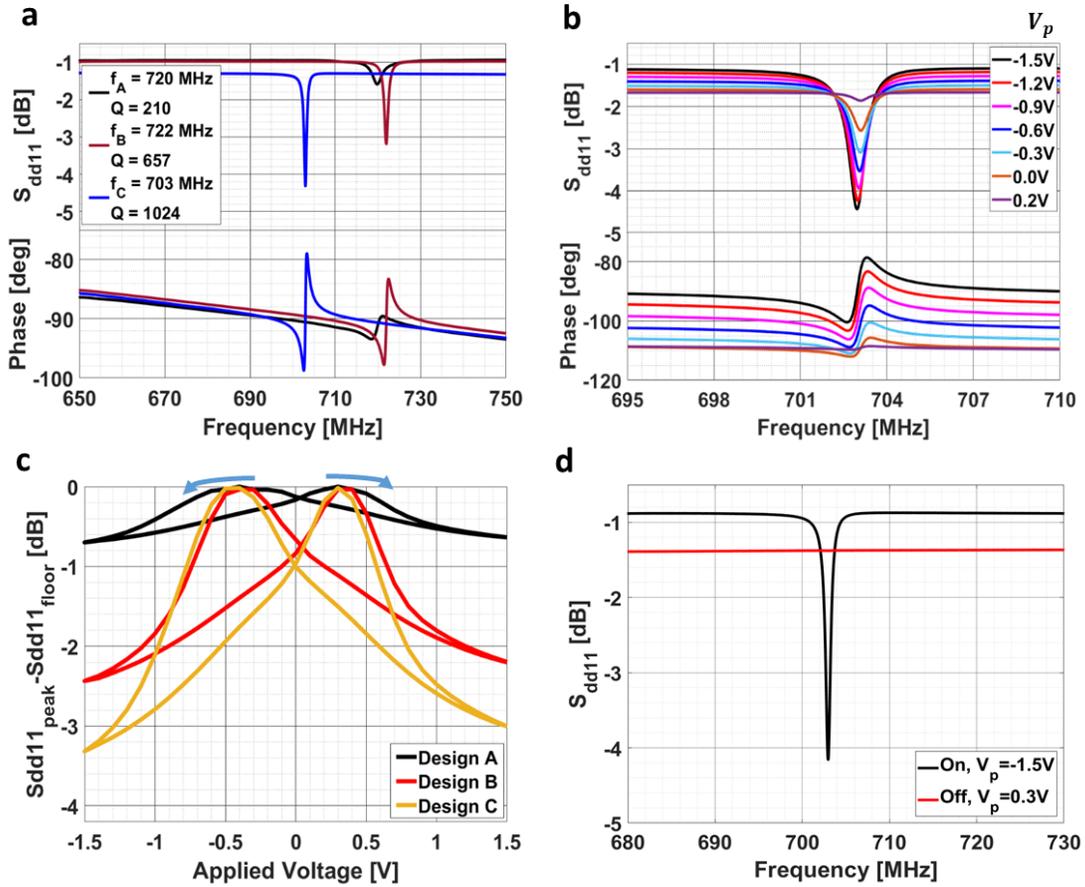

**Fig. 5 RF characterization of the FeCAP resonators. a** Comparison of $S_{21}$ for all the designs A-C **b** Measured frequency response of design C. **c** Due to the ferroelectric hysteresis, the $S_{21}$ magnitude demonstrates 'butterfly'-shaped variations with changing poling voltage. **d** At coercive voltages of ±0.3V, the FeCAP transducer can be completely switched off.

Figure 5 summarizes the measured RF response of the FeCAP resonators. We first consider resonator Design A, a waveguided resonance is observed at 720MHz with Q of 210 (black line in Fig. 5(a)). For Design B, the same resonance mode is found at 722 MHz but Q increased to 657 (brown line in Fig. 5(a)). Then, due to mass-loading from the added W volume, the resonance frequency for Design C shifts down to 703 MHz and Q increases to 1012. With an $f·Q$ product of $7.11\times10^{11}$ (blue line in Fig. 5(a)), the performance is 1.6x higher than the most state-of-the-art PZT resonators. In addition, the resonator is monolithically integrable with CMOS platform[10].

The dependence of $S_{dd11}$ on poling voltage is shown in Fig. 5(b) varying $V_p$ from -1.5V to 0.2V on both port 1(+) and port 1(−). The resulting 'butterfly'- shaped $S_{dd11}$ magnitude variation at resonance with poling voltage is shown in Fig. 5(c). The magnitude of $S_{dd11}$ reaches a minimum when all the transducers are biased at ±1.5V. Additionally, piezoelectric transduction is suppressed when the device is biased at the FeCAP coercive voltage of ±0.3V (Fig. 5(d)).

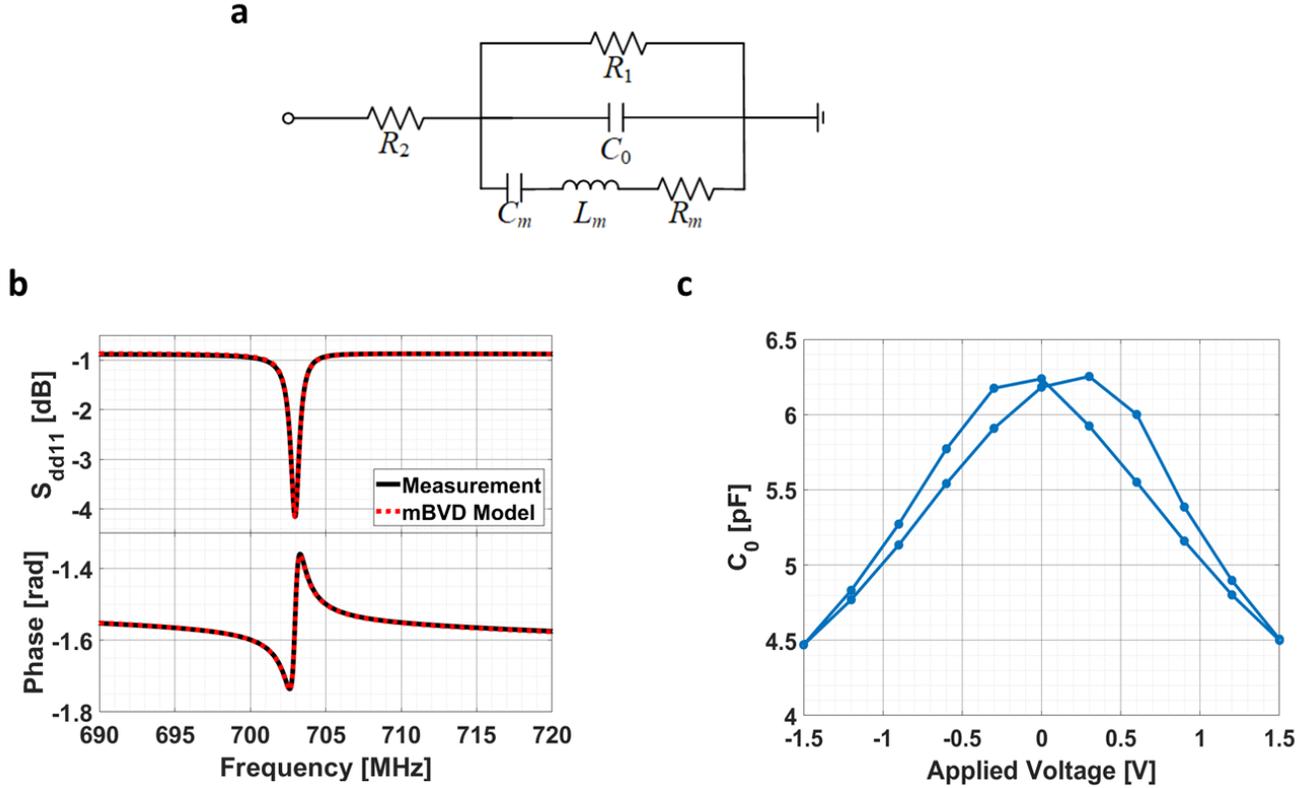

**Fig 6 a** Equivalent circuit of FeCAP resonator based on modified Butterworth-Van-Dyke model **b** $S_{dd11}$ of mBVD model and the measurement for the resonator with extensions and rectangular vias. **c** Measured $C_0$ variation with respect to the changing poling voltage.

A small signal equivalent circuit of the FeCAP resonator based on a modified Butterworth-Van-Dyke (mBVD) is presented in Fig. 6(a), consisting of a mechanical resonance branch which includes a motional resistance ($R_m$), motional inductance ($L_m$), and motional capacitance ($C_m$), in parallel with $C_0$. Here, $C_0$ is defined as the geometric capacitance of the structure which is valid for a fixed poling voltage. In the meantime, $R_1$ and $R_2$ model any resistive losses from leakage in the FeCAP and routing, respectively. The equivalent circuit parameters fitted to the measured data for design C under a poling voltage of -1.5V is shown in Figure 6(b). The corresponding circuit parameters is summarized in Table 1.

**Table 1 Performance of resonator Design C**

| Parameters | Value | Unit |
|---|---|---|
| $R_m$ | 134.5 | Ω |
| $L_m$ | 30.8 | µH |
| $C_m$ | 1.7 | fF |
| $C_0$ | 4.5 | pF |
| $R_1$ | 1959.8 | Ω |
| $R_2$ | 3.8 | Ω |
| $Q$ | 1012 | - |
| $k^2$ | 0.047 | - |

The electromechanical coupling coefficient can be extracted as:

$$k^2 = \frac{\pi^2}{8}\frac{C_m}{C_0} = 0.047\% \qquad (3)$$

Under the full bias sweep, $C_0$ varies between 4.6 pF and 6.5 pF as demonstrated in Fig. 6(c).

Thermal stability of resonance frequency was characterized for the FeCAP resonators over a temperature range from 23 to 90°C. The resonant frequency shift with respect to the temperature variation was summarized in ref. 31 and the Temperature Coefficient of Frequency was extracted as[31]:

$$TCF = \frac{1}{f_0}\frac{\partial f}{\partial T} = \text{-58.1±4.6 ppm/°C} \qquad (4)$$

This measured temperature sensitivity matches well to the predicted values based on the Temperature Coefficient of Young's Modulus ($TCE = \Delta E/\Delta T$) of the constituent materials in the CMOS BEOL[31].

## CONCLUSIONS

This work demonstrates a new class of ferroelectric-transduced RF MEMS resonators embedded seamlessly in CMOS, leveraging TI's 130nm FeRAM technology. The magnitude of the electromechanical response can be tuned by varying poling voltages on the transducers. Due to the hysteretic effect of the ferroelectric material, the magnitude of $S_{dd11}$ exhibit a hysteresis response with respect to changing poling voltages on the ports. The maximum Q of 1012 is obtained with optimization of vertical and lateral confinement of the acoustic mode in the CMOS stack. This corresponds to an $f·Q$ of $7.11\times10^{11}$.

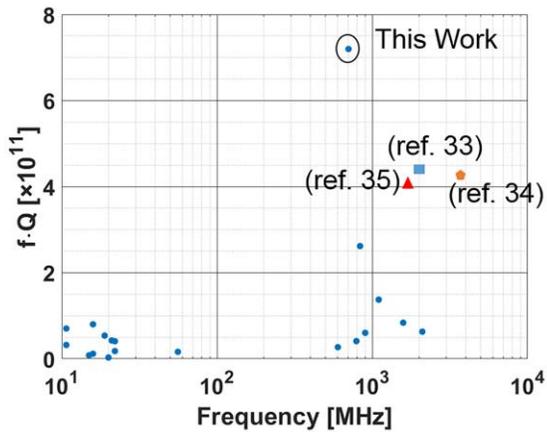

**Fig. 7** Comparison of *f·Q* with previous work

As indicated by Fig. 7, this performance is 1.6x higher than the state-of-the-art PZT resonators[32] with the additional benefit of CMOS integration. Suppression of electromechanical transduction is demonstrated when FeCAPs are biased at their coercive voltage of ±0.3V. The extracted TCF is -58.1±4.6 ppm/°C. Therefore, these devices provide a platform for applications include but not limited to RF components, timing, sensing, imaging, with CMOS integration.

## ACKNOWLEDGEMENT

The authors thank Texas Instruments for in-depth design and layout discussions and SEMs, as well as for fabrication of the devices presented here. This work was funded by the DARPA MTO UPSIDE program.

## AUTHOR CONTRIBUTIONS

Y.H. performed the COMSOL simulations, RF measurements, collected and analyzed experimental data, took the SEM image of the device and authored the manuscript. B.B. helped the layout and the fabrication of the devices. M.S and P.Y. performed the polarization-electric field measurements for the FeCAPs. D.W. supervised the project. All of the authors commented on the paper.

## CONFLICT OF INTEREST

The authors declare no conflict of interest.